# Potential of ChatGPT in predicting stock market trends based on Twitter Sentiment Analysis

Ummara Mumtaz & Summaya Mumtaz

**Abstract**

The rise of ChatGPT has brought a notable shift to the AI sector, with its exceptional conversational skills and deep grasp of language. Recognizing its value across different areas, our study investigates ChatGPT's capacity to predict stock market movements using only social media tweets and sentiment analysis. We aim to see if ChatGPT can tap into the vast sentiment data on platforms like Twitter to offer insightful predictions about stock trends. We focus on determining if a tweet has a positive, negative, or neutral effect on two big tech giants Microsoft and Google's stock value. Our findings highlight a positive link between ChatGPT's evaluations and the following day's stock results for both tech companies. This research enriches our view on ChatGPT's adaptability and emphasizes the growing importance of AI in shaping financial market forecasts.

## 1. Introduction

ChatGPT, developed by OpenAI, represents one of the latest advancements in the realm of artificial intelligence. It is grounded on the Generative Pre-trained Transformer (GPT) architecture (Vaswani et al., 2017), which enables it to understand and generate human-like text based on the input it receives. With its ability to engage in detailed, coherent, and contextually relevant conversations, ChatGPT has become a significant player in the AI industry (Gabashvili,2023). Its versatility allows it to be used across various domains - from casual chatbots to more complex tasks like content creation (email generation, text summarization, etc.) and data interpretation. However, their role in financial economics, especially in predicting stock market returns, is still a new area of exploration. While some might argue that these general-purpose models, not being specifically designed for stock prediction, may not be particularly useful, others believe that their vast training on extensive text data and ability to grasp natural language context could make them valuable for this purpose. The actual ability of LLMs in forecasting financial trends remains uncertain. Our research aims to address this by analyzing how well ChatGPT can use sentiment analysis to predict stock market returns.

In the digital age, the traditional means of predicting stock market trends—like analyzing quarterly reports or market fundamentals—have been complemented with more novel approaches, among which sentiment analysis stands prominent. Sentiment analysis harnesses the vast amount of unstructured data on the internet to gauge public sentiment, thereby offering insights into potential market movements. A particularly vibrant source of this sentiment data is Twitter, where millions of users express their opinions on a plethora of subjects, including the stock market, every day. ChatGPT, a state-of-the-art language model, presents a significant leap in processing and understanding such unstructured data. Its capability to comprehend context and nuance in textual content makes it a prime candidate for analyzing Twitter sentiments related to stock market trends.

But the challenge doesn't end at merely capturing sentiments; the ability to make predictions without being explicitly trained on specific tasks is a paramount advantage. This is where the zero-shot learning technique comes into play. Zero-shot learning (ZSL) allows models to make predictions or categorizations on data for which they haven't seen any examples during training (Larochelle et al., 2008; Lampert et al.,2014). In the context of machine learning, ZSL is often used in situations where labeled data for some tasks is scarce or unavailable. The integration of zero-shot learning in chatbot technologies like ChatGPT demonstrates an advancement that allows these models to respond accurately to a wide range of user prompts, even if they've never encountered them during training. In the context of stock market prediction using ChatGPT, it means that even if the model hasn't been explicitly trained on stock market data, it can leverage its generalized understanding of language to assess sentiments and make predictions on stock trends. This approach is not only cost-effective but also incredibly versatile, accommodating rapid shifts in market dynamics or unforeseen events which might not be well-represented in training data.

In this study, we explore the potential of using ChatGPT in predicting stock market trends solely based on Twitter sentiment analysis and by employing the zero-shot learning strategy. By enriching the potent capabilities of ChatGPT with the richness of sentiment data on Twitter, we aim to chart a novel path in stock market trend prediction and provide insights into the potential and limitations of such an approach. The rest of the paper is organized as: section 2 gives an overview of the existing literature; Section 3 describes the methodology including the data collection, pre-processing and temporal prediction modeling technique. Section 4 includes the conclusion and future directions.

## 2. Literature Review

Predicting stock market trends has been a topic of interest for decades, with traditional models primarily focusing on fundamental and technical analyses. However, with the advent of social media, researchers have explored the possibility of utilizing the vast amount of user-generated content as a predictive tool for stock market movements. We aim to consolidate key findings from various studies that have investigated the potential of using social media tweets, specifically from platforms like Twitter, to forecast stock market trends. Bollen et al. (2011) were among the pioneers to explore the relationship between Twitter sentiment and the stock market. They found a notable correlation between specific mood dimensions extracted from tweets and the Dow Jones Industrial Average. Zhang et al. (2011) also confirmed a significant relationship between Twitter sentiment and stock market movements but emphasized the importance of using sophisticated sentiment analysis tools. In recent years, advancements in machine learning have provided researchers with sophisticated tools. Nguyen et al. (2015) used Support Vector Machines (SVM) on Twitter data to predict stock prices and achieved a higher accuracy rate than traditional methods. Rao and Srivastava (2012) experimented with Naive Bayes classifiers, highlighting the importance of feature selection in sentiment analysis for accurate stock market prediction. While the sentiment of tweets is essential, the sheer volume of tweets mentioning a specific stock or related term can also be a predictor. Mao et al. (2012) found that the number of tweets about a company correlates positively with its trading volume. Siganos et al. (2014) argued for the advantages of real-time tweet analysis over daily aggregation, suggesting that intraday tweet volumes and sentiment shifts provide more immediate and actionable insights for traders.

---

[1] https://www.kaggle.com/datasets/khalidryder777/500k-chatgpt-tweets-jan-mar-2023

We found two relevant studies that recently tried to explore the potential of ChatGPT in stock market trend prediction. The research conducted by (Lopez-Lira et al., 2023) explored the capabilities of ChatGPT and other large language models in forecasting stock market returns based on news headlines. By classifying headlines as positive, negative, or neutral for stock prices using ChatGPT, a significant positive link between the model's scores and subsequent daily returns was identified. While ChatGPT surpasses conventional sentiment analysis techniques, basic models like GPT-1, GPT-2, and BERT lack precision in predicting returns. Notably, strategies leveraging ChatGPT-4 yield the highest Sharpe ratio. The study also reveals consistent underreaction in the market to company news, with stronger predictability in smaller stocks and those with negative news, hinting at the role of limits-to-arbitrage. The study by Xie and colleagues (Xie et al., 2023) found that ChatGPT, falls short in this financial context, lagging behind both cutting-edge and traditional forecasting methods, like linear regression. Even with advanced prompting strategies and incorporating tweets, the model's performance is lacking, further revealing issues in its explain ability and consistency. These findings underscore the potential need for model fine-tuning and pave the way for future research that combines social media sentiment with stock data to refine financial market predictions.

Not all studies found a consistent, strong correlation between Twitter sentiment and stock movements. Some, like Luss and d'Aspremont (2015), warned of potential overfitting when relying heavily on social media data, advocating for a mixed-methods approach. There's also the challenge of 'noise' in social media data. With the prevalence of bots and irrelevant content, filtering out noise remains a critical step in the analysis (Chen et al., 2018). The interplay between social media sentiment and stock market movements also raised questions about market efficiency and the potential for manipulation, as explored by researchers like Cook et al. (2018). Instead, our focus in this study is to evaluate whether ChatGPT, not trained in predicting returns, has the potential to predict stock market returns based only on tweets without considerable effort of cleaning the tweets and training or finetuning the model. Through a simple approach that leverages the model's stock market trend prediction capabilities, using sentiment analysis on tweets data and compare it to the actual stock market trends.

## 3. Methodology

### 3.1. Data Collection & Pre-processing :

We utilized a publicly available dataset at Kaggle named "500k ChatGPT-related Tweets Jan-Mar 2023"[1] for this analysis. The original dataset is composed of 500K tweets from January 4$^{th}$, 2023, to March 29$^{th}$, 2023, which were extracted by searching term "gpt" in the tweets. Tweets were specifically sought that mentioned or related to three distinct terms: 'ChatGPT', 'Microsoft', and 'Google'. This targeted approach ensured the relevance of the tweets to the topic of interest. We performed minimal data cleaning and only removed URL present in the tweets. Apart from URL removal, no other noise or extraneous data such as emotion icons or hashtags was filtered out, preserving the authentic voice and sentiment of the tweets.

Stock performance data for Google and Microsoft was procured from the NASDAQ website. As with the tweets, the data captured was between January 4$^{th}$, 2023 and February 28$^{th}$, 2023, ensuring synchronicity in the dataset's temporal scope. The stock market percentage change indicates the relative change in the value or price of a stock, index, or any other relevant market indicator over a specific period. It provides a standardized way to understand how much a stock, or the overall market, has increased or decreased, allowing for easy comparison over time or against other assets.

From the assembled tweet collection, samples were structured not based on the number of tweets but on the number of tokens. For each date, we picked a random sample consisting of a collection of tweets that cumulatively contained approximately 15,000 tokens. This method was chosen to standardize the volume of textual content processed by ChatGPT for each date and to ensure ChatGPT allowed token limit does not exceed.

### 3.2. Sentiment Analysis and Stock Market Prediction using ChatGPT

Instruction-Based Prompt Engineering Input: For sentiment analysis and trend prediction, ChatGPT was provided with a single consolidated input for each date. The input format was initiated with an instruction, outlining the task, followed by the tweets for that date. For instance, one example instruction is: "Based on the following tweets, predict the stock market trends for Google and Microsoft. We also provide instructions without using explicit company names as shown in the figure below.

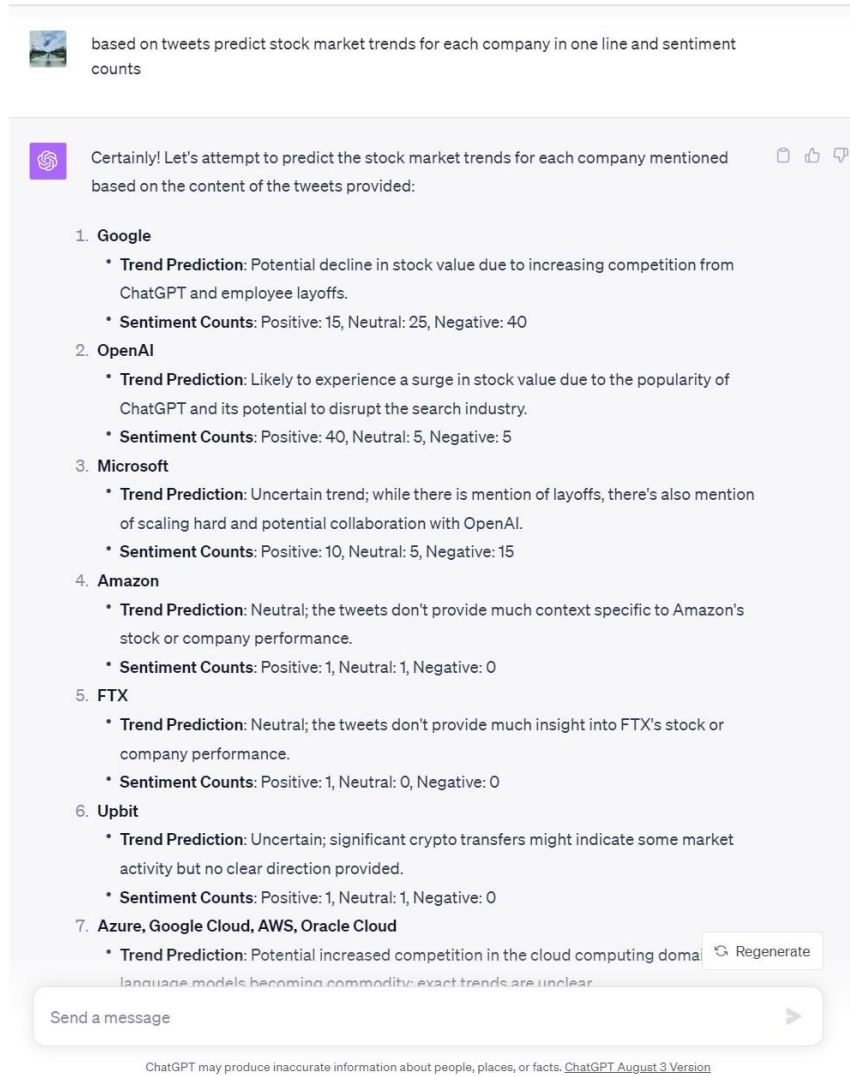

*Figure 1 : Example prompt 1 for stock market trend prediction and ChatGPT results*

ChatGPT was able to identify all the companies in the given tweet along with sentiment counts for each company. For each date, we collected the ChatGPT predictions in an excel file for both companies, along with the associated sentiment counts. We observed that without any fine tuning or few-shot learning, ChatGPT was able to bring relevant list of companies and the associated tweets count based on sentiments.

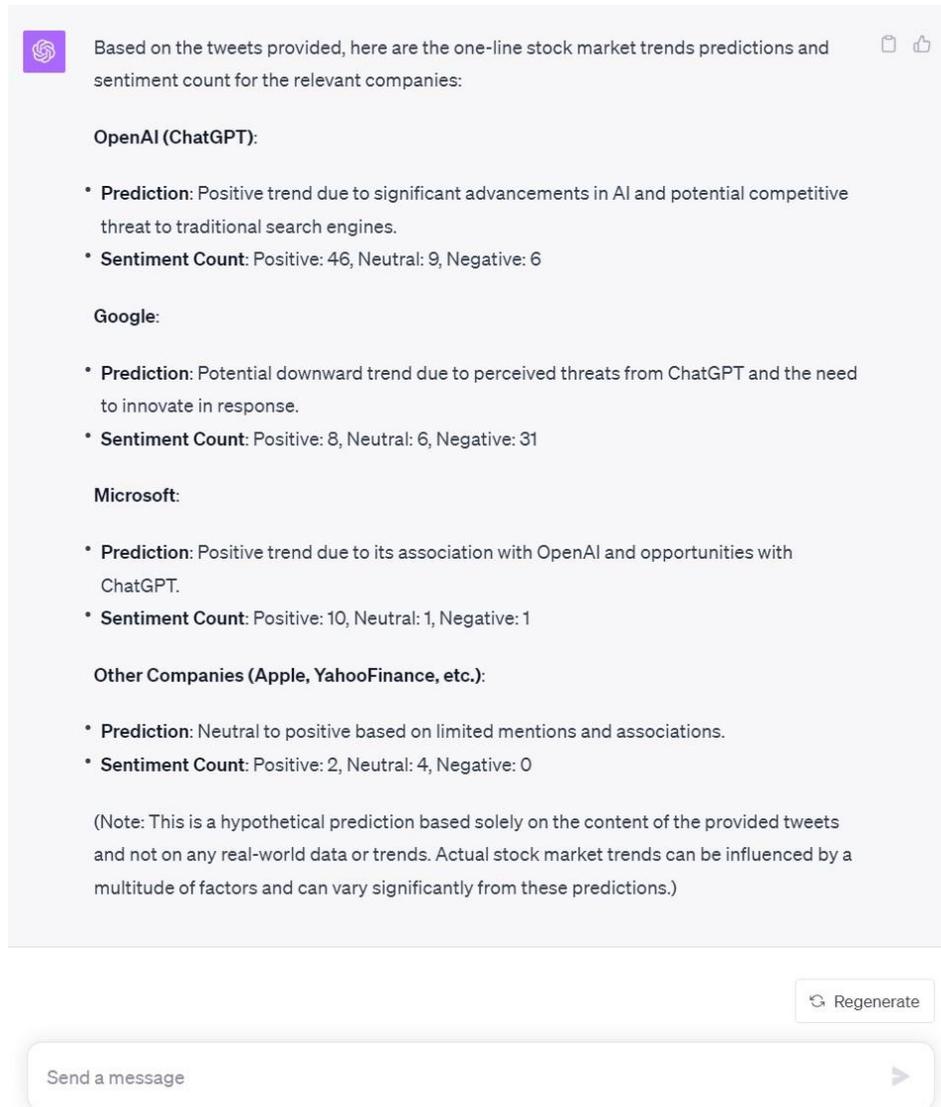

*Figure 2 : Example prompt 2 for stock market trend prediction and ChatGPT Results*

**3.3. Temporal Predictive Modelling**: We use temporal predictive modelling by utilizing a simple forward-looking approach. The stock market trend prediction from tweets on a specific day was utilized to predict stock market behavior for the subsequent day—before the market's opening. This was premised on the theory that today's public sentiment might influence tomorrow's stock performance. After obtaining ChatGPT's predictions for each date, these were systematically compared with the actual percentage changes in Google's and Microsoft's stock values as reported by NASDAQ. The objective was to ascertain the accuracy of the model's predictions and to determine the extent to which sentiment analysis could be used as a predictive tool for stock market trends. The model predicts the stock market trend in one to two sentences which cannot be directly compared with the percentage change in shares. To compare the results, we used a human evaluation approach and manually matched the stock market trend predicted by ChatGPT, with the stock percentage change of both companies on the following dates. Table 1. Below shows the data organization for evaluating the results.

*Table 1: Data Organization for Result evaluation*

| Date | % Change in Share | ChatGPT Prediction | Human Evaluation |
|---|---|---|---|
| 01/13/2023 | 1.086 | Stable | True |
| 01/17/2023 | -0.901 | Downward | True |
| 01/18/2023 | -0.186 | Positive | False |
| 01/19/2023 | 2.118 | Slight Bearish | False |
| 01/20/2023 | 5.341 | Decrease | False |
| 01/23/2023 | 1.806 | Bearish : Static trend | True |
| 01/24/2023 | -2.094 | Mixed : Potential Bearish | True |
| 01/25/2023 | -2.538 | Potential downward | True |

### 3.4. Results

To provide a more intuitive understanding of the results, we plot the ChatGPT predictions along with the actual % Change on the same date to visualize the results. Figure 3. shows the predictions made for Microsoft. The x-axis shows the prediction made on a specific date and y-axis refers to actual % change on the same date. For Microsoft, out of a 37-day period, ChatGPT's predictions aligned with the actual outcomes on 26 occasions. Thus, getting an accuracy of 70% for Microsoft stock market trend prediction.

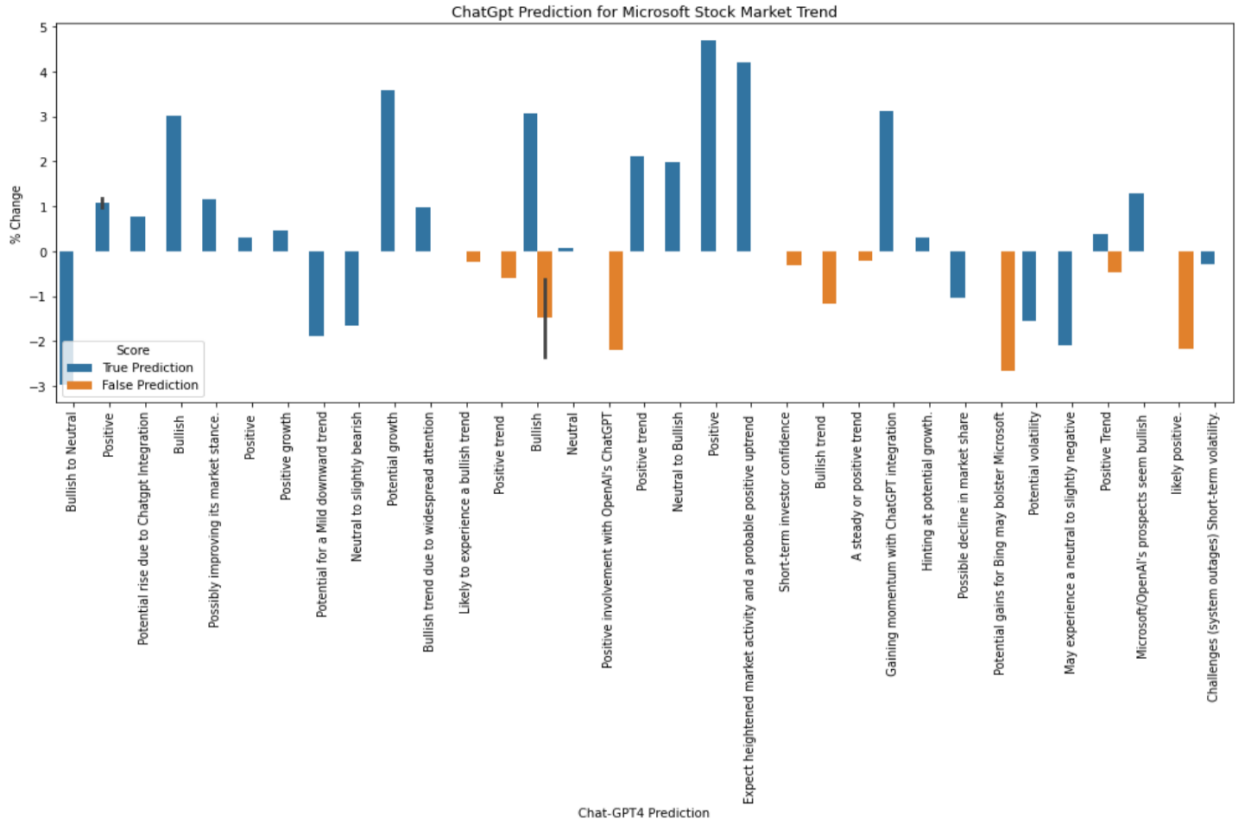

*Figure 3: ChatGPT Result Comparison for Microsoft with actual % Change (Blue color signifies correct predictions and orange color shows incorrect predictions). Bars are organized from left to right starting from January 5th, 2023, to February 28th, 2023.*

In assessing the predictive prowess of ChatGPT for Google's stock trends, it is found that the predictions deviated slightly from the actual outcomes. Specifically, out of a 36-day timeframe, the model made accurate forecasts on 23 days, resulting in an accuracy of roughly 63.88%. This accuracy, however, should be contextualized. Considering ChatGPT was not exclusively fine-tuned for stock market predictions and employing a zero-shot learning strategy, its performance is notably superior to a model making random predictions. Delving deeper, the model showcased an ability beyond mere trend prediction; it attempted to identify the underlying reasons for certain trends. As illustrated in "figure 4" concerning Google, ChatGPT not only made trend forecasts but also pinpointed specific factors like Google being "challenged by emerging competitors" or potential impacts from "Google's bard", and even identified a "potential threat from ChatGPT" itself. This reflects ChatGPT's capability to not just foresee the market direction but also recognize pivotal elements influencing that direction. Furthermore, it's crucial to emphasize that the analysis was conducted using a restrained set of tweets, hinting that a more exhaustive data set from Twitter or other social platforms might yield even better insights.

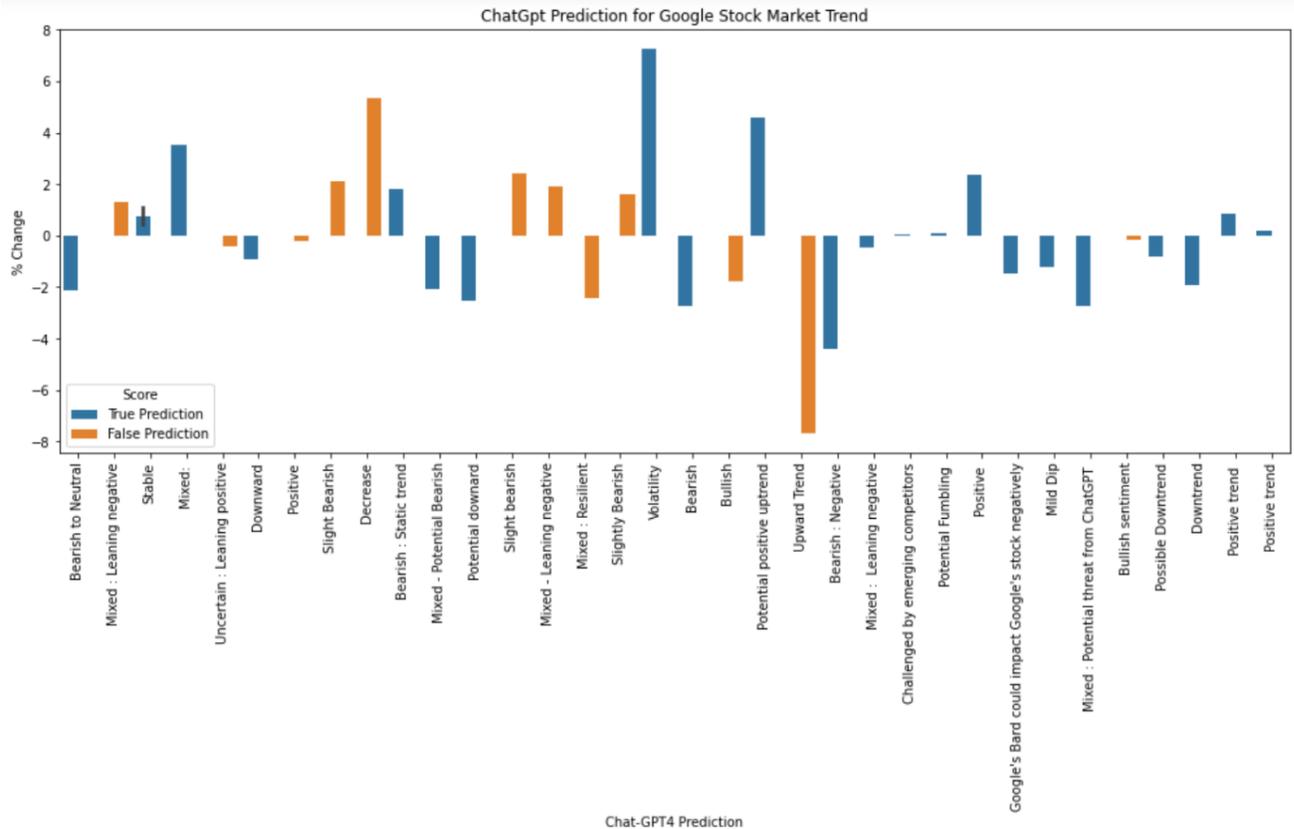

*Figure 4 : ChatGPT Result Comparison for Google with actual % Change (Blue color signifies correct predictions and orange color shows incorrect predictions). Bars are organized from left to right starting from January 5th, 2023, to February 28th, 2023.*

## 4. Conclusion and Future Work:

In this study, we've assessed ChatGPT's capability in predicting stock market trends for two major tech giants, Microsoft, and Google, utilizing only tweets and sentiment analysis. The results indicate that ChatGPT achieved an accuracy of 70% for Microsoft and 63.88% for Google. While the predictions weren't always spot-on, it's crucial to highlight that ChatGPT wasn't primarily designed for stock market predictions. Yet, its performance markedly exceeded that of a randomly predicting model. Further, its competency isn't confined to merely predicting trends. It also demonstrated its ability to identify underlying factors influencing those trends, adding depth and context to its forecasts. Given ChatGPT's encouraging performance, future endeavors could revolve around leveraging a more comprehensive dataset, possibly harnessing tweets, and data from other social media platforms to enhance predictive accuracy. It would be intriguing to observe how a richer dataset affects the model's forecasting capacity. Additionally, delving deeper into ChatGPT's ability to recognize and identify pivotal factors influencing market directions can be of immense value. It paves the way for more insightful and informed stock market predictions, which can be pivotal for traders and stakeholders.